\documentclass[aps,prb,twocolumn,superscriptaddress,showpacs]{revtex4-1}
\usepackage{amsmath,wasysym,amssymb,graphicx,epsfig,epstopdf,color,verbatim,ulem,braket,tabularx}
\usepackage[colorlinks,linkcolor=blue,citecolor=blue,urlcolor=blue]{hyperref}

\graphicspath{{Figures/}}

\begin{document}

\title{Combined spontaneous symmetry-breaking and symmetry-protected topological order from cluster charge interaction}

\author{Chen Peng}
\address{Department of Physics, Renmin University of China, Beijing 100872, China}
\author{Rong-Qiang He}
\email{rqhe@ruc.edu.en}
\address{Department of Physics, Renmin University of China, Beijing 100872, China}
\author{Yuan-Yao He}
\email{yhe@flatironinstitute.org}
\address{Center for Computational Quantum Physics, Flatiron Institute, New York, New York 10010, USA}
\address{Department of Physics, College of William and Mary, Williamsburg, Virginia 23187, USA}
\author{Zhong-Yi Lu}
\email{zlu@ruc.edu.en}
\address{Department of Physics, Renmin University of China, Beijing 100872, China}

\begin{abstract}
The study of symmetry-protected topological states in presence of electron correlations has recently aroused great interest as rich and exotic phenomena can emerge. Here, we report a concrete example by employing large-scale unbiased quantum Monte Carlo study of the Kane-Mele model with cluster charge interactions. The ground-state phase diagram for the model at half filling is established. Our simulation identifies the coexistence of a symmetry-protected topological order with a symmetry-breaking Kekul$\acute{e}$  valence bond order and shows that the spontaneous symmetry-breaking is accompanied by an interaction-driven topological phase transition (TPT). This TPT features appearance of zeros of single-particle Green's function and gap closing in spin channel rather than single-particle excitation spectrum, and thus has no mean-field correspondence. 
\end{abstract}

%\pacs{71.10.Fd, 02.70.Ss, 05.30.Rt., 11.30.Rd}

\date{\today}
\maketitle

\section{Introduction}
\label{sec:Intro}

The marriage of two ingredients, symmetry and topology, has greatly promoted the developments of topological phases of matter in the past decades~\cite{Hasan2010,XiaoLiang2011}. As one of the most important parts, topological insulators (TIs), defined as systems with gapped bulk spectrum and nontrivial gapless boundary spectrum, have been completely understood and classified for free fermion systems~\cite{Schnyder2009,Kitaev2009,Ryu2010}. When electron correlation steps in as the third ingredient, correlated symmetry-protected topological (SPT) phases~\cite{ChenXie2013} can exist as extensions of their noninteracting corresponding~\cite{Hohenadler2011,Zheng2011,Hohenadler2012,MENG2014} or even emerge from electron-electron interactions~\cite{Raghu2008,Sun2009,Yuanyao2018}. These SPT phases are characterized by various topological invariants~\cite{Kane2005a,Liang2006,Sheng2006,Prodan2009,Gurarie2011,Wang2012,Wang2013,Wang2014,Thomas2013,Hung2014,Yuanyao2016b,Yuanyao2016c} depending on symmetries, such as $Z_2$ invariant and spin Chern number. Electron-electron interactions can also drive interesting and exotic topological phase transitions (TPTs) in SPT states~\cite{Slagle2015,Yuanyao2016a,Yuanyao2016c,Wu2016,Qin2017}, which are characterized by changes of topological invariants and boundary spectrum instead of local order parameters. Thus, the TPTs between different SPT phases are generally beyond the standard Ginzburg-Landau (GL) phase transition paradigm. 

On the other hand, it's well known that electron-electron interactions can also be the driving force of various GL phase transitions and corresponding long-range ordered phases, which are characterized by spontaneous symmetry breaking and appearance of long-range orders. Representative examples are charge-density-wave~\cite{Huffman2014,WangLei2014,ZiXiang2015}, antiferromagetic~\cite{Assaad2013,Parisen2015,Otsuka2016}, and valence-bond-solid~\cite{WangDa2014} phases in fermion Hubbard models, spontaneously breaking inversion, spin rotation, and translation symmetries, respectively. These disordered-ordered quantum phase transitions can be either of first order or continuous, which, for the latter case, belong to certain universality class, depending on the dimension and symmetry. 

With these in mind, it will be of great interest to demystify whether {\it topological symmetry protection} and {\it spontaneous symmetry breaking} can coexist in a single quantum phase in fermion systems, namely an SPT phase with symmetry breaking and long-range order. Furthermore, the corresponding TPT and disordered-ordered phase transition in such a system should be quite different from the conventional ones mentioned above. Such exotic phases have already been studied in bosonic systems~\cite{Daniel2019}, termed as symmetry-breaking topological insulators, while the fermionic counterpart is missing.

In this paper, we present a positive answer to above question via a concrete example of interacting fermion model studied by numerically exact, large-scale quantum Monte Carlo (QMC) simulations. Our QMC results have revealed a topologically nontrivial Kekul$\acute{e}$ valence bond solid (KVBS) phase, which spontaneously breaks $Z_3$ symmetry (three ways to form a KVBS long-range order) and has spin Chern number $C_s=-1$. We have also found coinciding KVBS phase transition and TPT from $C_s=+1$ to $C_s=-1$ inside quantum spin-Hall insulator (QSHI) phases. Moreover, this TPT is accompanied by appearance of zeros of single-particle Green's function. The spin channel becomes critical while the single-particle gap remains finite across this exotic phase transition, which makes this TPT very different from the standard TPT in free fermion systems.

The rest of this paper is organized as follows. In Sec.~\ref{sec:modelmethod}, we first present the interacting fermion model we studied and the QMC algorithm we employed. The physical quantities calculated in this work is also briefly reviewed. Then, the QMC simulation results, as the main part of this work, are presented and discussed in details in Sec.~\ref{sec:Results}. Finally, Sec.~\ref{sec:SumDiscuss} summarizes this work, and discusses the possible extensions in future works.

\section{Model and method}
\label{sec:modelmethod}

\subsection{Kane-Mele Model with cluster charge interaction}

Our model describes a topological insulator of fermions with cluster charge interaction on a honeycomb lattice. The Hamiltonian contains tight-binding and interaction parts $\hat{H}=\hat{H}_0 + \hat{H}_U$ as
\begin{eqnarray}
\label{eq:model}
\hat{H} =&& -t\sum_{\langle\mathbf{i},\mathbf{j}\rangle\alpha}(c_{\mathbf{i}\alpha}^+c_{\mathbf{j}\alpha} + c_{\mathbf{j}\alpha}^+c_{\mathbf{i}\alpha}) \\ \nonumber
  && + i\lambda\sum_{\langle\langle \mathbf{i},\mathbf{j} \rangle\rangle\alpha\beta}\nu_{\mathbf{ij}}(c_{\mathbf{i}\alpha}^+\sigma_{\alpha\beta}^z c_{\mathbf{j}\beta} - c_{\mathbf{j}\beta}^+\sigma_{\beta\alpha}^z c_{\mathbf{i}\alpha} ) \\ \nonumber
  && + U\sum_{\hexagon}(\hat{Q}_{\hexagon}-6)^2,
\end{eqnarray}
where $\mathbf{i},\mathbf{j}$ represent the lattice sites, $\alpha,\beta=\uparrow,\downarrow$ label fermion spins, and $\lambda$ is the strength of spin-orbit coupling. $\hat{H}_0$ is the Kane-Mele model, consisting of the nearest-neighbor (NN) hopping and the intrinsic spin-orbit coupling (SOC), and the factor $\nu_{\mathbf{ij}}=-\nu_{\mathbf{ji}}=\pm1$ depends on the orientation of the next-nearest-neighbor bonds as demontrated in Fig.~\ref{fig:LatPhDigm}(a). $\hat{H}_U$ stands for the cluster charge interaction in which the summation runs over all the hexagons on honeycomb lattice. $\hat{Q}_{\hexagon}=\sum_{\mathbf{i}\in\hexagon}\hat{n}_{\mathbf{i}}$ is the total charge operator in a hexagon with $\hat{n}_{\mathbf{i}}=\sum_{\alpha}c_{\mathbf{i}\alpha}^+c_{\mathbf{i}\alpha}$. Throughout this work, we set $t$ as the energy unit for simplicity. 

The competition between the nontrivial band topology and the electron correlation in our model in Eq.~(\ref{eq:model}) is quite explicit. First of all, the Kane-Mele model has a QSHI ground state with counterpropagating edge states~\cite{Kane2005a,Kane2005b}. This QSHI phase is an SPT phase protected by $U(1)_{\rm spin}\times U(1)_{\rm charge}\rtimes Z_2^T$ (with $Z_2^T$ as the time-reversal) symmetry, which results in $\mathbb{Z}$ classification. The appropriate topological invariant for describing the QSHI phase is the spin Chen number $C_s=(C_{\uparrow}-C_{\downarrow})/2$ with $C_{\uparrow}$ and $C_{\downarrow}$ as the Chern numbers in spin-up and down channels~\cite{Yuanyao2016b,Yuanyao2016c}, respectively. Furthermore, this QSHI phase is stable against weak and local interactions, and thus it survives in small $U$ region in our model. As for the cluster charge interaction, it has been shown~\cite{XiaoYan2018} that it favors a KVBS and antiferromagnetic insulating phases at intermediate and strong interactions, respectively, when the SOC term in Eq.~(\ref{eq:model}) is turned off. In the thermodynamic limit, the KVBS phase breaks the $Z_3$ symmetry, and forms the pattern of alternating strong and weak bonds like the one (of three) shown in the inset of Fig.~\ref{fig:LatPhDigm}(c). In this work, we only concentrate on the KVBS phase induced by intermediate cluster charge interaction. It's explicit that the broken symmetry in the KVBS phase and the symmetry protecting the QSHI phase in Kane-Mele model are in different symmetry sectors. This means that these two phases may coexist in the phase diagram, which is the key point that we engaged to demystify in this paper by employing numerically unbiased large-scale QMC simulations. 

\subsection{Projector quantum Monte Carlo method}

We apply the projector QMC (PQMC) method, the zero-temperature version of the determinantal QMC algorithm, to study the ground-state properties of the model in Eq.~(\ref{eq:model}) at half-filling, i.e., one electron per site on average. The PQMC algorithm calculates the ground-state expectation results of static (equal-time) observables within the projected wavefunction as 
\begin{equation}
\label{eq:GSObs}
\langle {\hat O} \rangle  = \lim_{\Theta\to+\infty}\frac{{\left\langle {{\Psi _T}} \right|{e^{-\Theta \hat{H}/2}}\hat O{e^{-\Theta \hat{H}/2}}\left| {{\Psi _T}} \right\rangle }}{{\left\langle {{\Psi _T}} \right|{e^{ - \Theta \hat{H}}}\left| {{\Psi _T}} \right\rangle }},
\end{equation}
where $|\Psi_T\rangle$ is a trial wavefunction nonorthogonal to the true ground state of the many-body system. With a large enough but finite projection parameter $\Theta$, the many-body ground state as $|\Psi_g\rangle = e^{-\Theta\hat{H}/2}|\Psi_T\rangle$ can be achieved for a finite-size system. Three steps need to be done for carrying out the imaginary-time projection $e^{-\Theta\hat{H}/2}|\Psi_T\rangle$ in Eq.~(\ref{eq:GSObs}) by the PQMC method. First, we divide the projection parameter into $M$ slices as $\Theta=M\Delta_{\tau}$ and $e^{-\Theta\hat{H}}=(e^{-\Delta_{\tau}\hat{H}})^M$, where $\Delta_{\tau}$ needs to be small enough. Second, we apply the Trotter decomposition, $e^{-\Delta_{\tau}\hat{H}}=e^{-\Delta_{\tau}\hat{H}_0}e^{-\Delta_{\tau}\hat{H}_U}+\mathcal{O}(\Delta_{\tau}^2)$, to separate the free fermion and interaction  parts in the interacting model. The Trotter error $\mathcal{O}(\Delta_{\tau}^2)$ in this step is fully controlled. Third, we decouple the interaction term into free fermions coupled to auxiliary fields by Hubbard-Stratonovich (HS) transformation. After that, standard importance sampling like the Metropolis algorithm can be performed to the auxiliary-field configurations and physical observables can be computed through single-particle Green's function. Similar formula like Eq.~(\ref{eq:GSObs}) exists for dynamic quantities, including the imaginary-time correlation function in both fermionic and bosonic channels.

For the model we studied in Eq.~(\ref{eq:model}), we adopt the HS transformation with four-component Ising fields to decouple the cluster charge interaction~\cite{Assaad2005}. Fixed at half-filling, the model in Eq.~(\ref{eq:model}) is free of minus sign problem as we can prove that the weight of every single auxiliary-field configuration are nonnegative due to the particle-hole symmetry. In this work, we choose $\Delta_{\tau}t=0.05$ and $\Theta t \ge 80$ for  the linear system sizes we simulated as $L=9, 12, 15, 18, 21$ (with number of lattice sites $N_s=2L^2$). These choices of $\Delta_{\tau}$ and $\Theta$ parameters have been tested to fully get rid of the Trotter error and converge to the true many-body ground state, respectively. 

\begin{figure}[t]
\includegraphics[width=0.98\columnwidth]{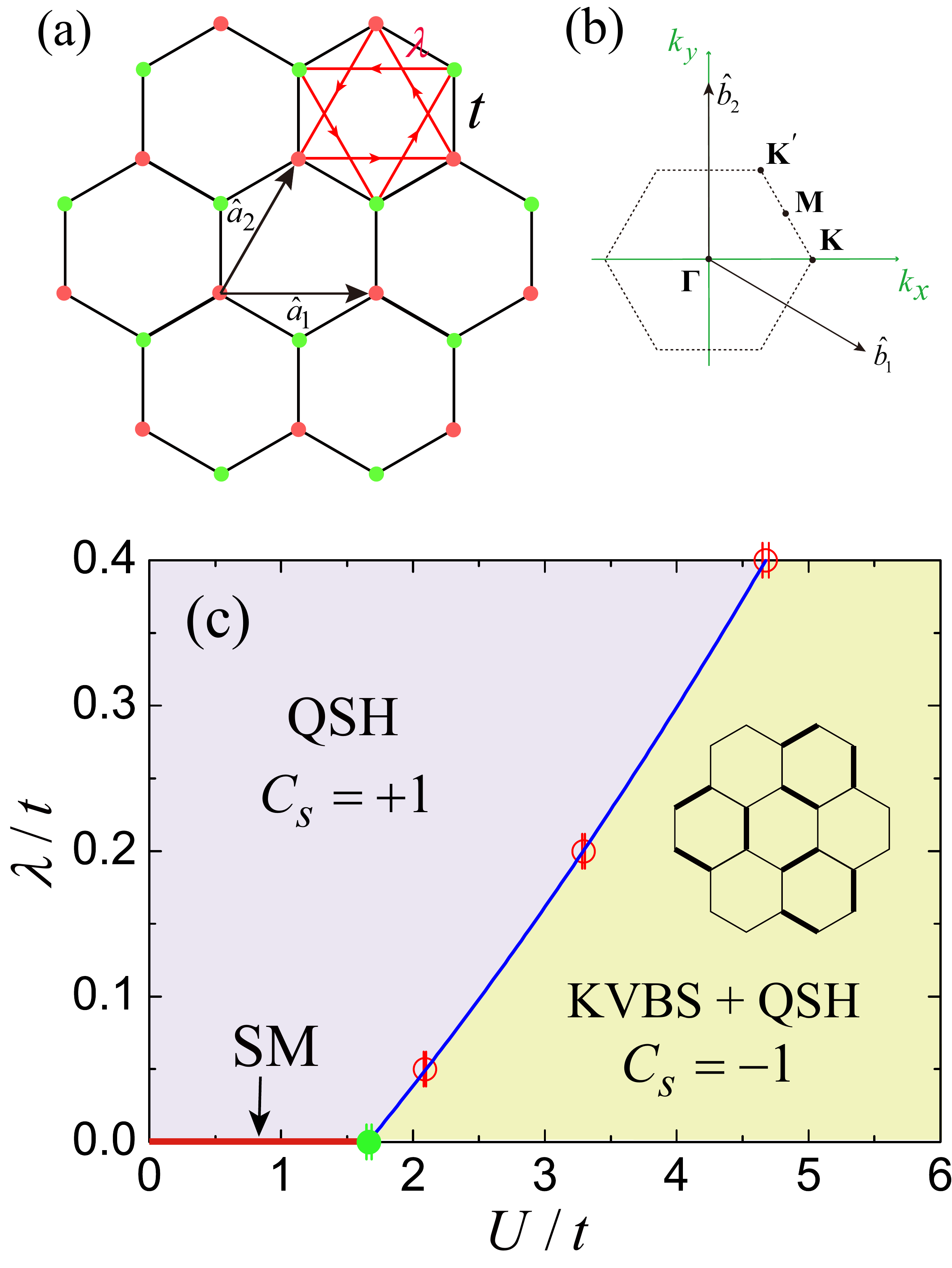}
\caption{\label{fig:LatPhDigm} The lattice structure and ground-state $\lambda$-$U$ phase diagram. (a) Illustration of the honeycomb lattice and Hamiltonian (\ref{eq:model}). A unit cell of the lattice, indicated by $\hat{a}_1$, $\hat{a}_2$, contains A (red dot) and B (green dot) sublattices. The black lines represents NN hopping, and the red lines shows the SOC with arrows indicating $\nu_{\mathbf{ij}}=+1$. (b) Brillouin zone of the model and the reciprocal lattice vector $\hat{b}_1$ and $\hat{b}_2$ are given corresponding to $\hat{a}_1$ and $\hat{a}_2$, respectively. (c) The $\lambda$-$U$ phase diagram from PQMC simulations. Three cases of $\lambda/t=0.05,0.2,$ and $0.4$ are studied. The disordered-KVBS ordered phase transition is verified to coincide with the TPT from $C_s=+1$ to $C_s=-1$, as shown by the solid, blue line. Thus, a topologically nontrivial KVBS phase is established. As a comparison, the quantum phase transition with $U_c/t=1.666(8)$ for $\lambda=0$ from Dirac semimetal to a trivial KVBS phase from Ref.~\onlinecite{XiaoYan2018} is also indicated by the solid, green dot. }
\end{figure}

We have measured various physical observables to determine the ground-state phase diagram of model (\ref{eq:model}). We first detect the formation of long-range KVBS order by measuring the correlation functions and its structure factor in reciprocal space as
\begin{equation}
\label{eq:KVBSOrder}
C_{\rm VBS}(\mathbf{k})=\frac{1}{L^2}\sum_{\mathbf{mn}}e^{i\mathbf{k}\cdot(\mathbf{R}_{\mathbf{m}}-\mathbf{R}_{\mathbf{n}})}\langle\hat{B}_{\mathbf{m}}\hat{B}_{\mathbf{n}}\rangle,
\end{equation}
where $\hat{B}_{\mathbf{m}}=\sum_{\alpha}(c_{\mathbf{m}\alpha}^+c_{\mathbf{m}+\boldsymbol{\delta}\alpha}+h.c.)$ with $\mathbf{m,n}$ as indices for unit cells and $\boldsymbol{\delta}$ standing for one of the three NN directions. The corresponding ordering vector for the KVBS order is $\mathbf{K}$ and $\mathbf{K}^{\prime}$ in the Brillouin zone (shown in Fig.~\ref{fig:LatPhDigm}(b)). Thus, the order parameter of the KVBS can be defined as $\Delta_{\mathbf{K}}=\sum_{\mathbf{m}}e^{i\mathbf{K}\cdot\mathbf{R}_{\mathbf{m}}}\langle\hat{B}_{\mathbf{m}}\rangle$, where the phase factor $e^{i\mathbf{K}\cdot\mathbf{R}_{\mathbf{m}}}$ signifies the $Z_3$ symmetry breaking. Then the location of the quantum phase transition between the disordered and KVBS ordered phases can be determined via the correlation ratio $R_{\rm corr}=1-\frac{C_{\rm VBS}(\mathbf{K}+\mathbf{q})}{C_{\rm VBS}(\mathbf{K})}$ where $\mathbf{q}$ represents the smallest momentum in the Brillouin zone for the corresponding finite-size lattice. The histogram of the KVBS order around the phase transition is also computed to better characterize the transition. 

To obtain the information of excitations of the system, we measure the dynamic single-particle Green's function and spin-spin correlation functions as 
\begin{eqnarray}
\label{eq:Dynamic}
G(\mathbf{k},\tau)&&=\frac{1}{4L^2}\sum_{\mathbf{mn},\gamma\alpha}e^{i\mathbf{k}\cdot(\mathbf{R}_m-\mathbf{R}_n)}\langle c_{\mathbf{m}\gamma,\alpha}^+(\tau)c_{\mathbf{n}\gamma,\alpha}(0)\rangle, \hspace{0.5cm} \\ \nonumber
S(\mathbf{k},\tau)&&=\frac{1}{2L^2}\sum_{\mathbf{mn},\gamma}e^{i\mathbf{k}\cdot(\mathbf{R}_m-\mathbf{R}_n)}\langle \hat{s}_{\mathbf{m}\gamma}^z(\tau)\hat{s}_{\mathbf{n}\gamma}^z(\tau) \rangle,
\end{eqnarray}
where $\gamma=A,B$ for the sublattices. Then the single-particle gap $\Delta_{sp}(\mathbf{k})$ and spin gap $\Delta_{s}(\mathbf{k})$ can be extrapolated from the large-$\tau$ behavior of $G(\mathbf{k},\tau)$ and $S(\mathbf{k},\tau)$ as $G(\mathbf{k},\tau)\propto e^{-\Delta_{sp}(\mathbf{k})\tau}$ and $S(\mathbf{k},\tau)\propto e^{-\Delta_{s}(\mathbf{k})\tau}$, respectively. For model (\ref{eq:model}), the global minimum of single-particle gap is either at $\mathbf{k}=\mathbf{K} (\mathbf{K}^{\prime})$ or $\mathbf{k}=\mathbf{M}$ depending on the SOC strength $\lambda$, while the minimal spin gap is at $\mathbf{k}=\boldsymbol{\Gamma}$.

The TPTs driven by interactions can be dramatically different from those in the free fermion systems. To characterize the topological nature of the phases and quantum phase transition for model (\ref{eq:model}), we employ the technique of computing spin Chern number via zero-frequency single-particle Green's function~\cite{Wang2012,Wang2013,Wang2014}. All the details of this calculation are carefully demonstrated and tested in Ref.~\onlinecite{Yuanyao2016b}. For model (\ref{eq:model}), the time-reversal symmetry guarantees $C_{\uparrow}=-C_{\downarrow}$, and thus we have $C_s=C_{\uparrow}$.

\section{Numerical results and discussions}
\label{sec:Results}

\subsection{Ground-state phase diagram}
\label{sec:PhaseDiagram}

We first briefly summarize the $\lambda$-$U$ phase diagram obtained by our PQMC simulations for model (\ref{eq:model}), as shown in Fig.~\ref{fig:LatPhDigm}(c). With finite $\lambda$, the disordered QSHI phase with $C_s=+1$ at $U=0$ extends to weak but finite interaction regime as expected. Strikingly, a QSHI phase with $C_s=-1$ coexisting with the long-range KVBS order is discovered at intermediate interactions. Furthermore, the TPT from $C_s=+1$ to $C_s=-1$, which is accompanied by the presence of zeros of single-particle Green's function, coincides with the disordered-ordered phase transition characterized by the appearance of the long-range KVBS order breaking $Z_3$ symmetry. Our simulation results also suggest that the quantum phase transition between these two phases is of first order (at least for $\lambda/t>=0.2$). Across the phase transition point, the excitation gap in fermionic channel (single-particle gap) remains finite while the spin gap closes and reopens. These interesting behaviors render the interaction-driven TPT in our model rather exotic and dramatically different from that in free fermion system, where pole of single-particle Green's function appears and single-particle gap vanishes. As a comparison, a quantum phase transition between the Dirac semimetal and the KVBS phase with emergent $U(1)$ symmetry at the transition point has been established~\cite{XiaoYan2018} for the $\lambda=0$ case, which is also shown in Fig.~\ref{fig:LatPhDigm}(c). Thus, the presence of SOC alters the ground states and the corresponding phase transitions to a great extent. Besides, we have also confirmed the antiferromagnetic Mott insulator phase is absent in the interaction range $U/t=0\sim6$, for arbitrary $\lambda/t$.

In this work, we have carried out PQMC simulations for $\lambda/t=0.05,0.2,0.4$ as presented in Fig.~\ref{fig:LatPhDigm}(c). In the following sections, we mainly demonstrate the PQMC results for $\lambda/t=0.05$ and $0.2$ cases as the only difference for $\lambda/t=0.4$ case is the net shift in phase boundary. 

\subsection{The KVBS phase transitions}
\label{sec:DisOrdTrans}

\begin{figure}[t]
\includegraphics[width=0.99\columnwidth]{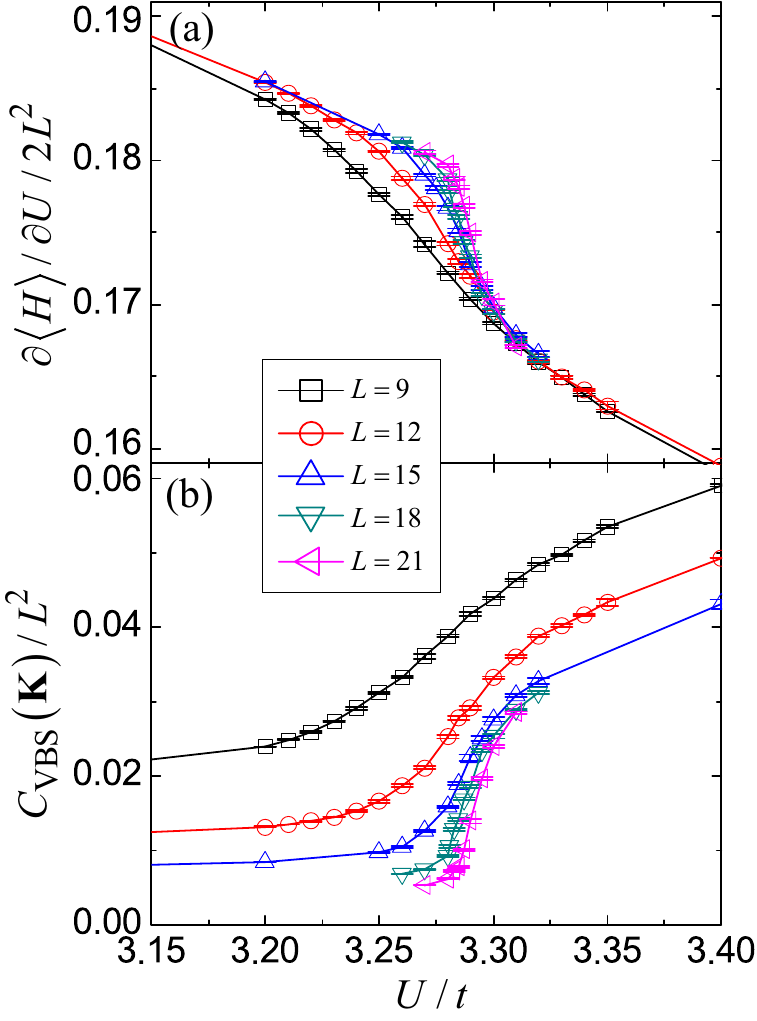}
\caption{\label{fig:EnergyOrder} (a) The total energy density derivative $\partial\langle\hat{H}\rangle/\partial U/2L^2$, and (b) the structure factor $C_{\rm VBS}(\mathbf{K})/L^2$ of KVBS for $\lambda/t=0.2$ with $U/t$ crossing the quantum phase transition. The results for linear system sizes with $L=9,12,15,18$, and $21$ are shown.}
\end{figure}

\begin{figure}[t]
\includegraphics[width=0.99\columnwidth]{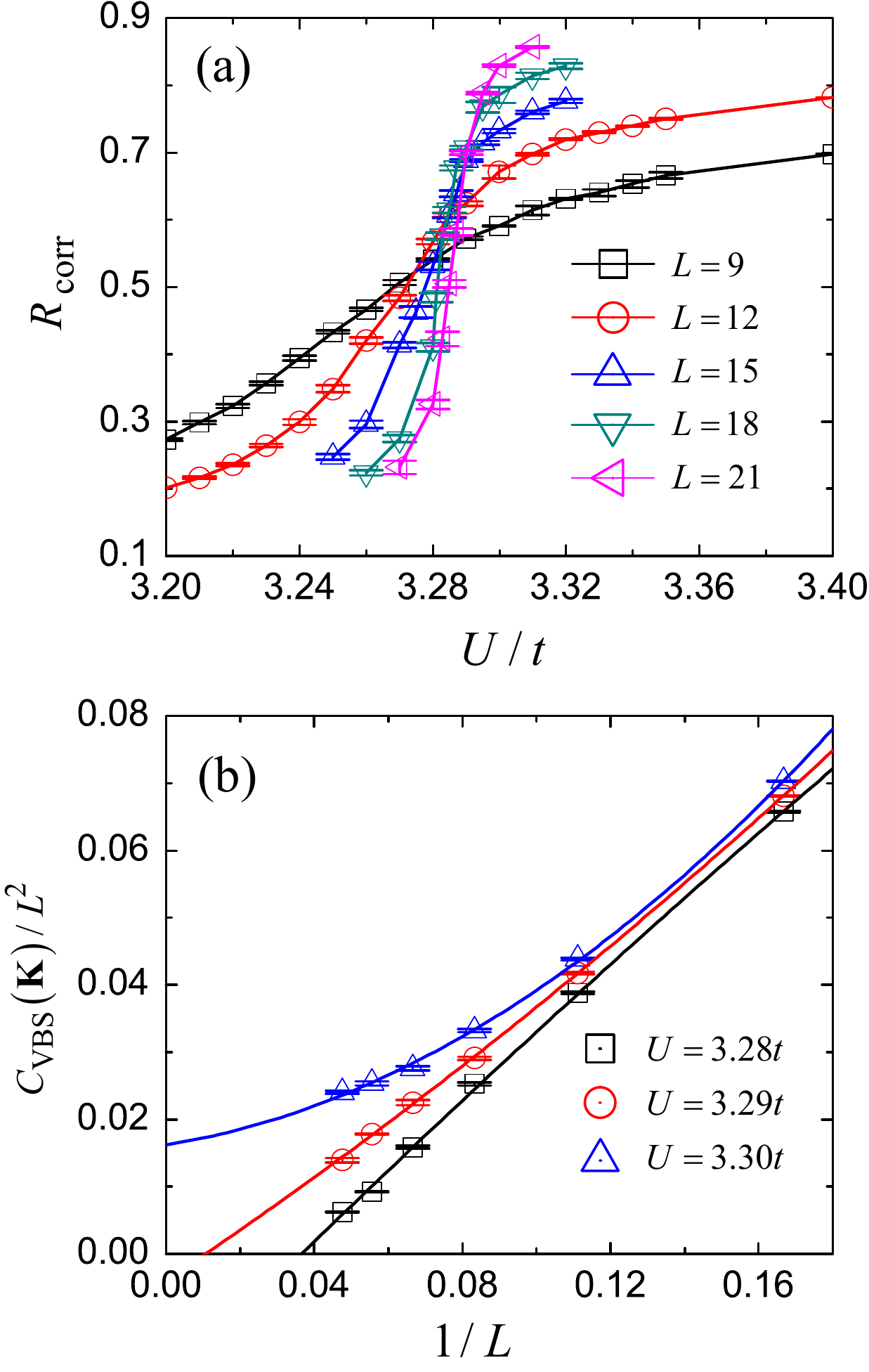}
\caption{\label{fig:CorrRatioOrdFiniteSize} (a) The correlation ratio $R_{\rm corr}$ for the KVBS order across the phase transition, and (b) finite-size scaling of structure factors of KVBS order with $U/t=3.28,3.29,$ and $3.30$ for $\lambda/t=0.2$.}
\end{figure}

The disordered-KVBS ordered phase transition, characterized by the formation of long-range KVBS order and breaking $Z_3$ symmetry, is first determined from the energies and the bond-bond correlation functions defined in Eq.~(\ref{eq:KVBSOrder}). 

The results for the total energy density derivative $\partial\langle\hat{H}\rangle/\partial U/2L^2$ and the structure factor $C_{\rm VBS}(\mathbf{K})/L^2$ for $\lambda/t=0.2$ with increasing $U/t$ are presented in Fig.~\ref{fig:EnergyOrder}. The explicit kinks in the results of the total energy density derivative around $U/t=3.30$ indicate the location of the quantum phase transition. Similar kinks can also be observed in the results of the structure factor. An important observation is that with increasing system size, the kinks in all these results tend to evolve into jumps. This can be taken as the indication of first-order disordered-ordered phase transition, as the magnitude of the KVBS order can be evaluated as $m_{\rm VBS}=\sqrt{C_{\rm VBS}(\mathbf{K})/L^2}$. Besides, the finite-size effect turns out to be quite strong for $U/t<3.30$ region, and all the results shown in Fig.~\ref{fig:EnergyOrder} converge within $L=21$ for $U/t>3.32$. 

\begin{figure}[t]
\includegraphics[width=0.99\columnwidth]{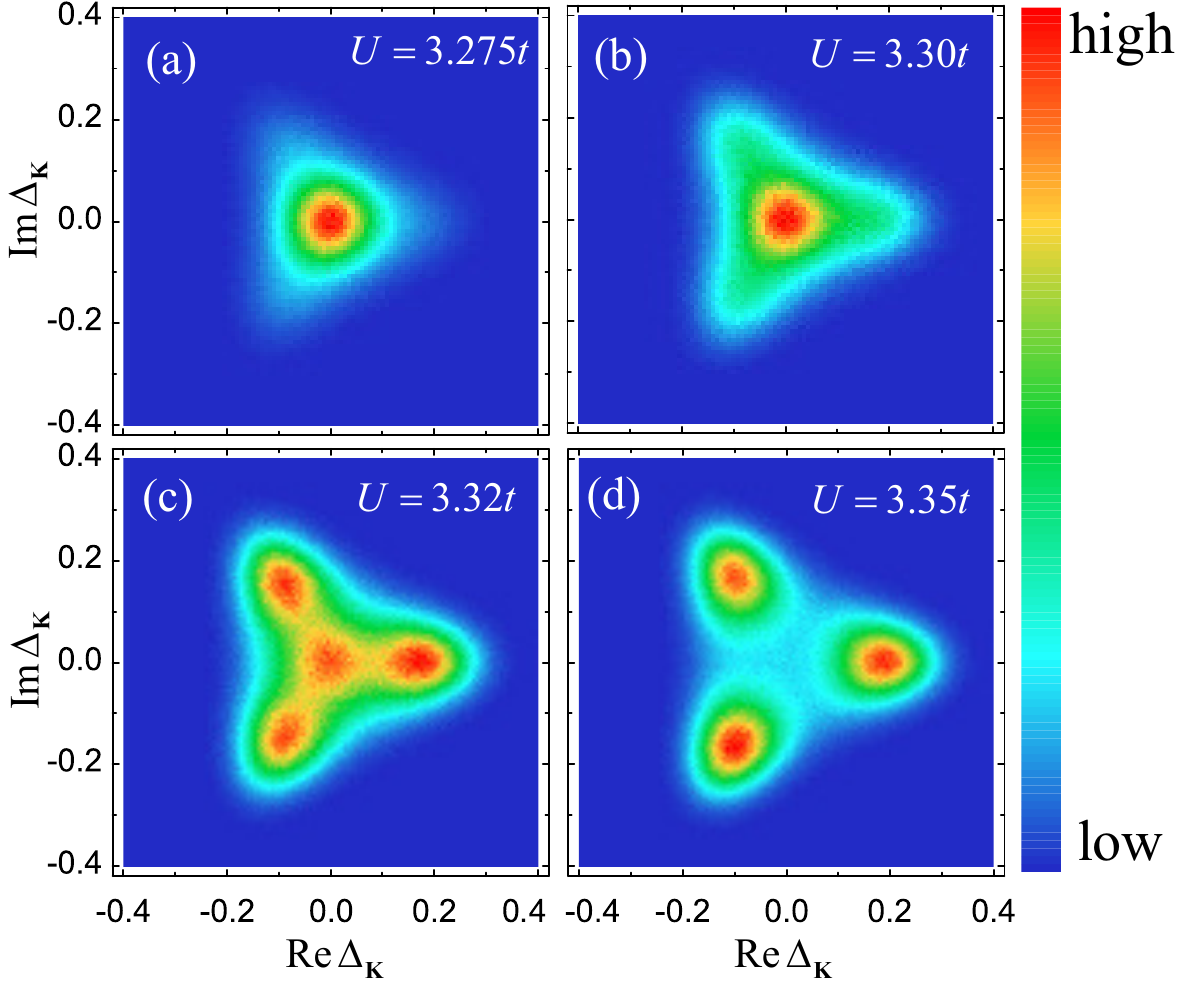}
\caption{\label{fig:Histogram} Histogram of the KVBS order parameter $\Delta_{\mathbf{K}}$ for $\lambda/t=0.2$ across the phase transition with $L=12$. $U/t=3.275$, $U/t=3.30$, $U/t=3.32$ and $U/t=3.35$ (a-d).}
\end{figure}

To explicitly determine the location of this phase transition, Fig.~\ref{fig:CorrRatioOrdFiniteSize} shows the results of correlation ratio $R_{\rm corr}$ and finite-size scaling for KVBS order parameter around the phase transition for $\lambda/t=0.2$. As we know, $R_{\rm corr}$ saturates to identity and vanishes in the ordered and disordered phases, respectively, in the thermodynamic limit. The crossings of $R_{\rm corr}$ results of systems with different sizes can be taken as the phase transition point~\cite{Yuanyao2018}. As shown in Fig.~\ref{fig:CorrRatioOrdFiniteSize}(a), though the finite-size effect is quite significant, we can observe the convergence of the crossings from $U/t\simeq3.27$ for $L=9$ and $L=12$ to $U/t\simeq3.29$ for $L=18$ and $L=21$. Similar to the results shown in Fig.~\ref{fig:EnergyOrder}, the smooth curve of $R_{\rm corr}$ also tends to change into jump with increasing system size, which is also an indication of first-order phase transition. A careful finite-size scaling of KVBS order parameters within $U/t=3.28,3.29,3.30$ are shown in Fig.~\ref{fig:CorrRatioOrdFiniteSize}(b). These results explicitly show that in the thermodynamic limit, there is no long-range KVBS order for $U/t=3.28,3.29$, while $m_{\rm VBS}=0.12(2)$ can be extrapolated for $U/t=3.30$. This dramatic change in $m_{\rm VBS}$ within such small variation of $U/t$ parameter also suggests a first-order phase transition. We have also tried the data collapse for results of structure factors for KVBS order and no reliable results can be extracted, which contradicts with the hypothesis of continuous phase transition. From the results in Fig.~\ref{fig:CorrRatioOrdFiniteSize}, we can extract the location of the phase transition as $U_c/t=3.295(5)$ for $\lambda/t=0.2$ case. 

In Fig.~\ref{fig:Histogram}, the histogram of the KVBS order parameter $\Delta_{\mathbf{K}}$ for $\lambda/t=0.2$ across the phase transition is shown with $L=12$. Before the transition ($U/t=3.275$), $\Delta_{\mathbf{K}}$ completely distributes around zero indicating absence long-range order. As a comparison, $\Delta_{\mathbf{K}}$ mainly resides in three different patches after the transition, which represents three types of ordering connected by $Z_3$ symmetry (and the system falls into one of them in the thermodynamic limit as the symmetry breaking). The most important observation is the coexistence of the peak around zero and the other three peaks centered around finite values with $U/t=3.32$, which is actually the fingerprint of first-order phase transition. Due to the finite-size effect, the value of $U/t$, where this coexistence is observed, is slightly different from the transition point in the thermodynamic limit. 

All the above results are well consistent and suggest that the disordered-KVBS ordered phase transition is of first order. We obtained similar results for $\lambda/t=0.4$ with $U_c/t=4.652(5)$.

Here we also present the understanding of the above quantum phase transition from theoretical aspect. The Landau cubic criterion shows that the phase transitions cannot be continuous if Ginzburg-Landau (GL) free energy contains cubic terms of order parameters. This is indeed the case of the above KVBS phase with $Z_3$ symmetry breaking, where the cubic terms is generally allowed in GL free energy. In Ref.~\onlinecite{ZiXiang2017}, it was shown that such kind of disordered-ordered quantum phase transition can actually be continuous due to the coupling between the KVBS order parameter and gapless Dirac fermions, namely the fermion-induced quantum critical points. This is the case for $\lambda=0$ as shown in the phase diagram in Fig.~\ref{fig:LatPhDigm}(c). With finite $\lambda$, the fermions are gapped out (as shown in Sec.~\ref{sec:FermiBoseGaps}) and Landau cubic criterion applies. Thus, this can explain the first-order quantum phase transition we observed above.

\subsection{The topological phase transitions}
\label{sec:TopoTrans}

To characterize the topological nature of the quantum states in the phase diagram in Fig.~\ref{fig:LatPhDigm}(c), we calculate spin Chern number $C_s$ from $\mathbf{G}_{\alpha}(i\omega=0,\mathbf{k})$ (with $\alpha=\uparrow,\downarrow$), and we also examine the eigenvalues of $\mathbf{G}_{\alpha}(0,\mathbf{k})$ to get more information of the structure of poles and zeros of the single-particle Green's function across the TPT.

\begin{figure}[t]
\includegraphics[width=0.99\columnwidth]{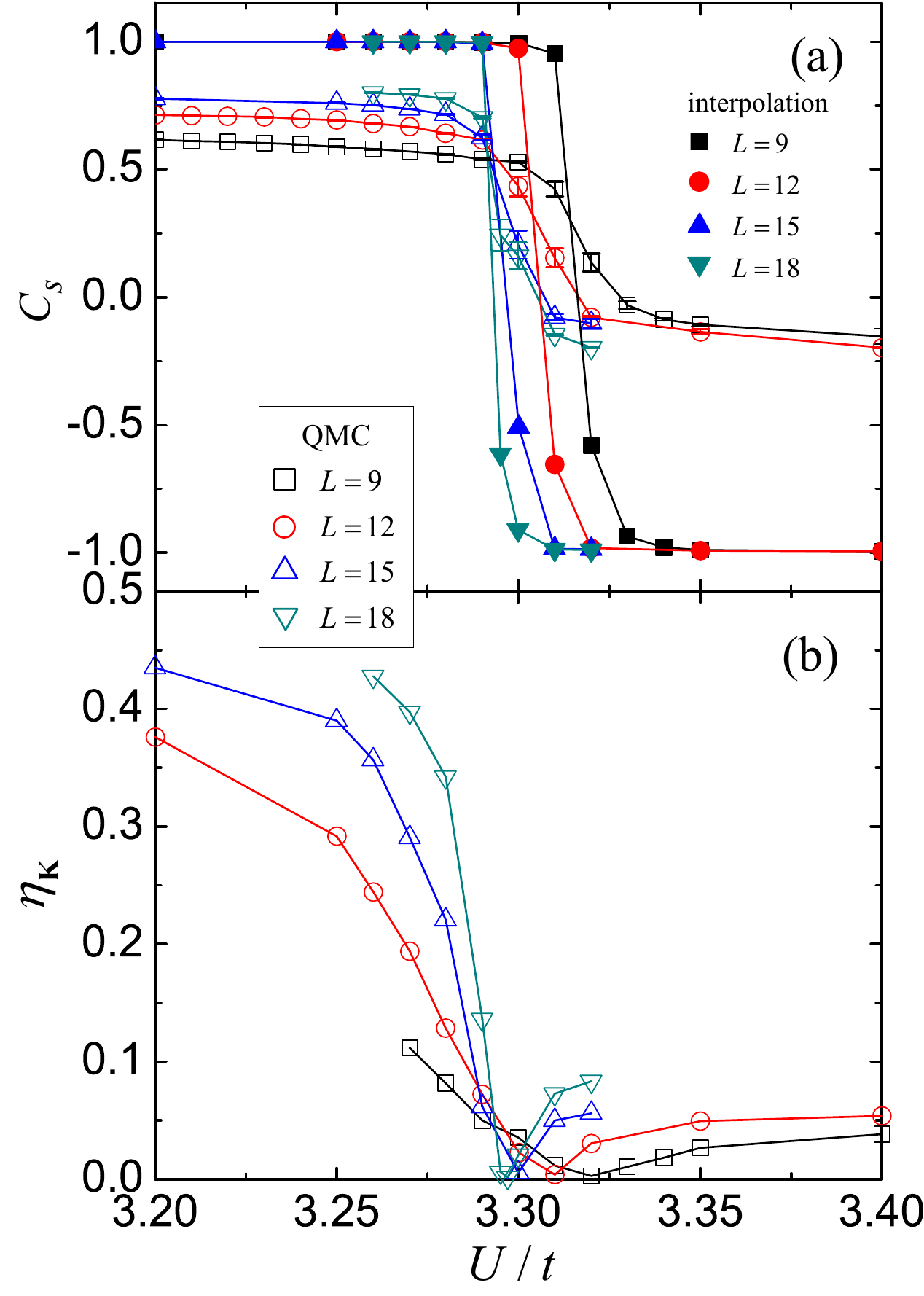}
\caption{\label{fig:TopInvtEigVal} (a) The spin Chern number $C_s$, and (b) positive eigenvalue $\eta_{\bf{K}}$ of $\mathbf{G}_{\sigma}(i\omega=0,\mathbf{K})$ matrix across the TPT for $\lambda/t=0.2$. In (a), both results of direct QMC calculations in finite-size systems and application of the interpolation technique (see Ref.~\onlinecite{Yuanyao2016b}) are shown.}
\end{figure}

Figures.~\ref{fig:TopInvtEigVal}(a) and (b) show the results of $C_s$ and the positive eigenvalue of $\mathbf{G}_{\uparrow}(0,\mathbf{K})$ matrix for $\lambda/t=0.2$ across the TPT. As expected, the spin Chern number obtained from direct, finite-size PQMC calculations is smooth across the transition and it's far from integer quantized values. After the interpolation calculations~\cite{Yuanyao2016b}, we indeed observe integer jumps from $C_s=+1$ to $C_s=-1$ across the TPT. Again, the finite-size effect is quite significant for the first several system sizes, and the best estimate of the location of TPT is $U_c/t=3.29(1)$, which is well consistent with the critical value obtained for disordered-KVBS ordered phase transition in Sec.~\ref{sec:DisOrdTrans}. Thus, these two transitions coincide. More importantly, we can conclude that the KVBS phase in the phase diagram in Fig.~\ref{fig:LatPhDigm}(c) is also a QSHI phase with $C_s=-1$, combining the results in Sec.~\ref{sec:DisOrdTrans} and Fig.~\ref{fig:TopInvtEigVal}(a). On the other hand, the spin Chern number changes only because of appearance of poles or zeros in the single-particle Green's function. We have indeed observed the zeros of $\mathbf{G}_{\uparrow}(0,\mathbf{k})$ at both $\mathbf{K}$ and $-\mathbf{K}$ points at the transition point from the positive eigenvalue of $\mathbf{G}_{\uparrow}(0,\mathbf{K})$, as shown in Fig.~\ref{fig:TopInvtEigVal}(b) . This also explains the change of spin Chern number $\Delta C_s=2$ across the TPT, since there are two zeros in $\mathbf{G}_{\uparrow}(0,\mathbf{k})$ at $\mathbf{K}$ and $-\mathbf{K}$. In addition, we have also confirmed the absence of poles in single-particle Green's function, which is consistent with the fact that the single-particle gap remains open across the TPT as shown in Sec.~\ref{sec:FermiBoseGaps}. Similar results are obtained for $\lambda/t=0.05$, as presented and discussed in Appendix~\ref{sec:ResultOf005}. 

The topologically nontrivial KVBS phase as well as the exotic TPT discussed above is quite different from those in free fermion (or mean-field) systems in several aspects. We have also studied the corresponding mean-field system by replacing the cluster charge interaction $\hat{H}_U$ in Eq.~(\ref{eq:model}) by the term $m_{\rm VBS}\sum_{\langle\mathbf{ij}\rangle\alpha}(c_{\mathbf{i}\alpha}^+c_{\mathbf{j}\alpha}+h.c.)$ with finite and zero $m_{\rm VBS}$ for the strong and weak bonds, respectively, as shown in the set of Fig.~\ref{fig:LatPhDigm}(c). The results are presented in Appendix~\ref{sec:MeanField}. With increasing $m_{\rm VBS}$, the QSHI phase with $C_s=+1$ evolves into a topologically trivial phase with $C_s=0$ at a finite value of $m_{\rm VBS}$. And the TPT in this mean-field system is accompanied by single-particle gap closing and reopening. Thus, the appearance of the KVBS phase with $C_s=-1$ in model (\ref{eq:model}) originates from electron correlations. Theoretically, the nontrivial KVBS phase should change into a trivial KVBS phase if $m_{\rm VBS}$ in the thermodynamic limit is large enough. However, the antiferromagnetic Mott insulator, which breaks the spin $U(1)$ symmetry and time-reversal symmetry, takes over for strong interactions before the KVBS order completely breaks the nontrivial topology. This is consistent with our numerical results that the trivial KVBS phase is absent in the phase diagram.

\subsection{Excitation gaps}
\label{sec:FermiBoseGaps}

\begin{figure}[t]
\includegraphics[width=0.99\columnwidth]{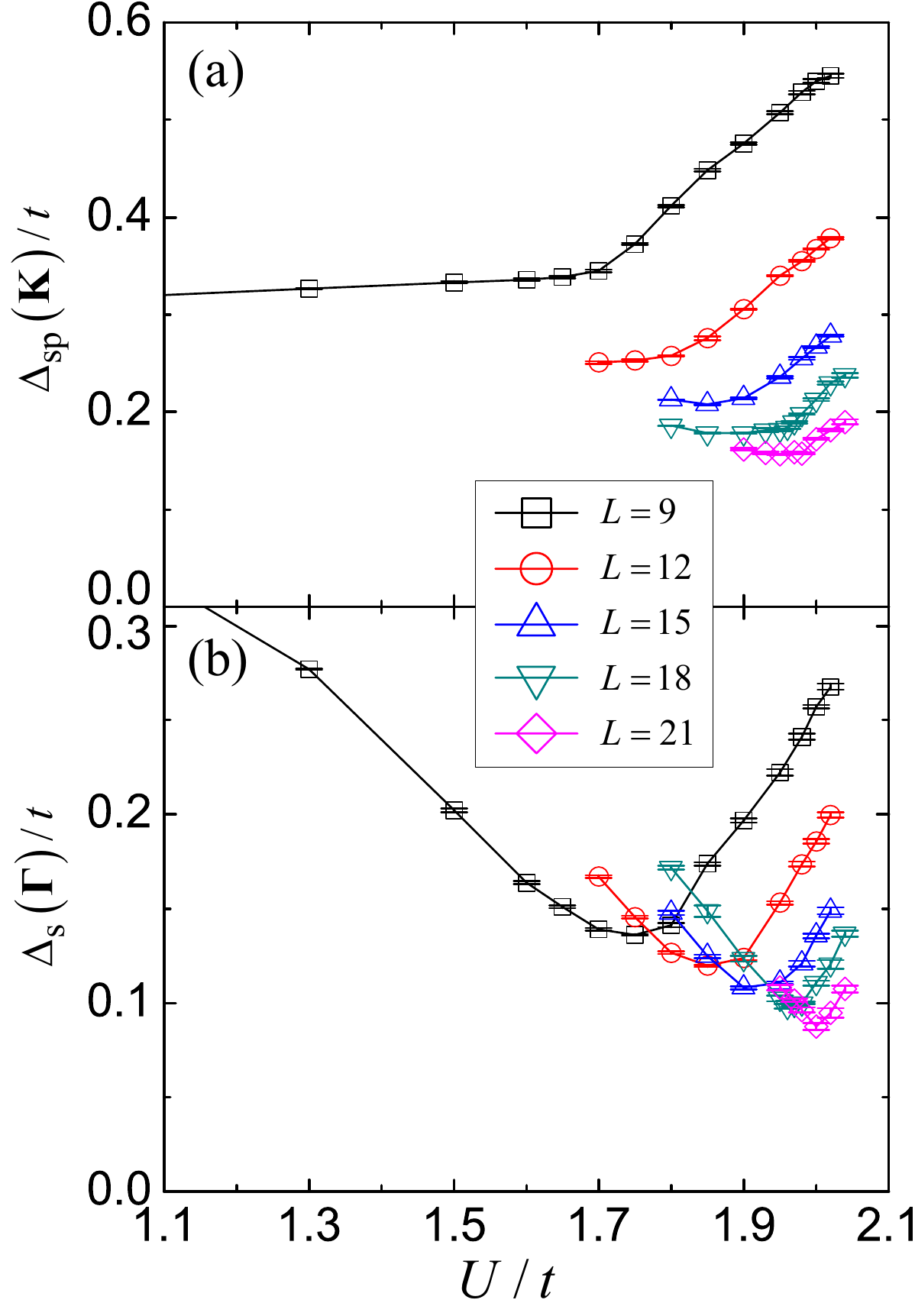}
\caption{\label{fig:FermiBoseGap} (a) The single-particle gap $\Delta_{sp}(\mathbf{K})/t$, and (b) the spin gap $\Delta_{s}(\boldsymbol{\Gamma})/t$ for $\lambda/t=0.05$ around the phase transition $U_c/t=2.09(1)$. }
\end{figure}

\begin{figure}[t]
\includegraphics[width=0.99\columnwidth]{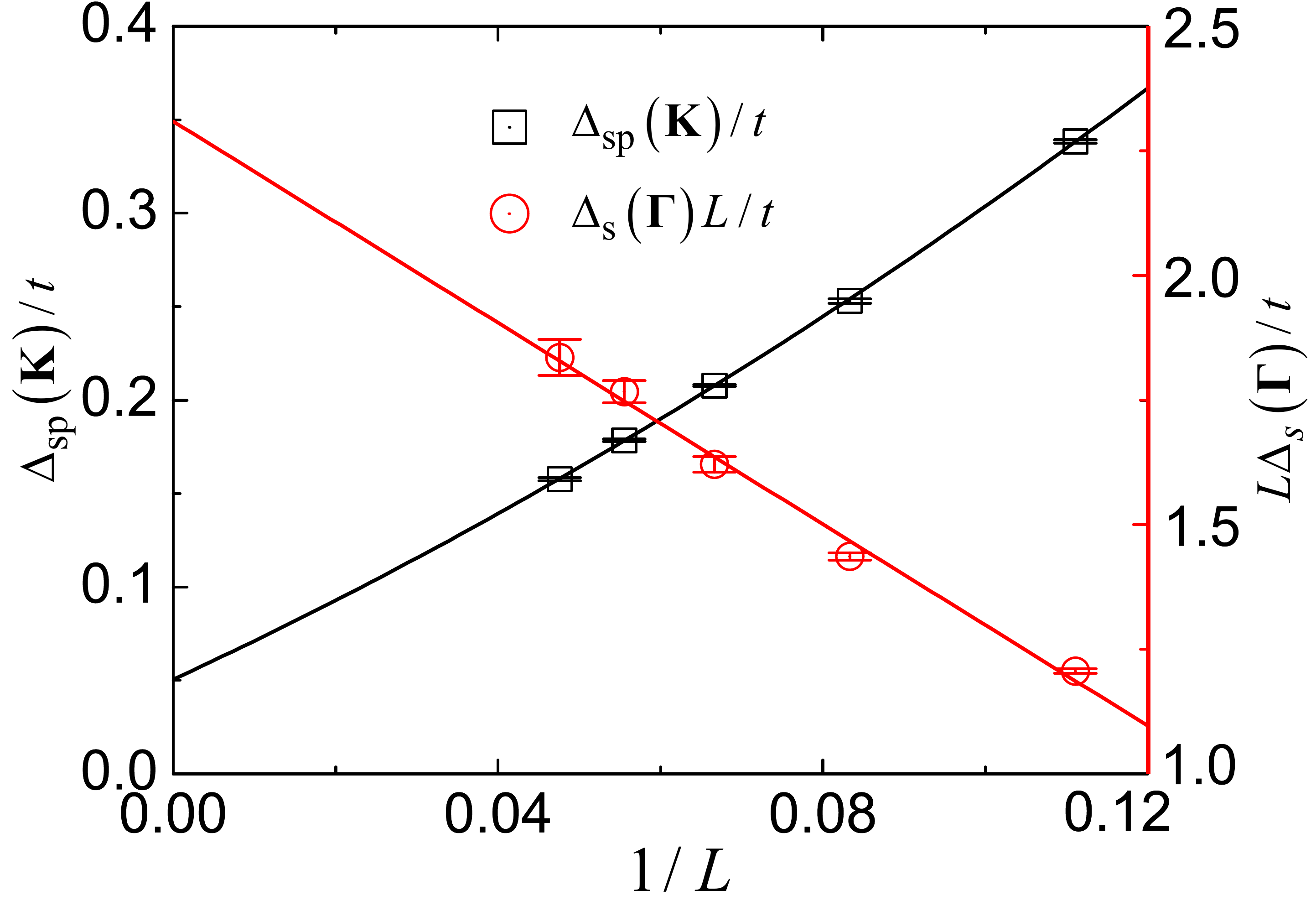}
\caption{\label{fig:GapsExtrapolate} Finite-size scaling of the single-particle gap $\Delta_{sp}(\mathbf{K})/t$ and the spin gap $L\Delta_{s}(\boldsymbol{\Gamma})/t$ with the dip values of different system sizes in Fig.~\ref{fig:FermiBoseGap} for $\lambda/t=0.05$. A quadratic and linear fitting are applied for $\Delta_{sp}(\mathbf{K})/t$ and $L\Delta_{s}(\boldsymbol{\Gamma})/t$, respectively.}
\end{figure}

Both the single-particle gap and the spin gap for $\lambda/t=0.05$ around the quantum phase transition are presented in Fig.~\ref{fig:FermiBoseGap}. In Appendix~\ref{sec:RawdataDynamic}, the raw data of imaginary-time single-particle Green's function and spin-spin correlation function defined in Eq.~(\ref{eq:Dynamic}) is also presented and discussed. There is a general dip in the curves of both gap when varying parameter $U/t$ for all the system sizes. The location of the dips also changes with increasing system size because of finite-size effect. The spin gap is significantly smaller than the single-particle gap around the transition point, highlighting the fact that the low-energy excitations of this interacting system are indeed bosonic as mentioned in Sec.~\ref{sec:DisOrdTrans}. In Fig.~\ref{fig:FermiBoseGap}(b), the value of spin gap is even larger in the system with larger system size at some specific parameter $U/t$ before the transition. This effect has also been observed in the Kane-Mele-Hubbard model with small SOC parameter~\cite{Hohenadler2012}.  

To reliably extrapolate the excitation gaps at the transition point and in the thermodynamic limit, we carry out the finite-size scaling of $\Delta_{sp}/t$ and $L\Delta_s/t$ with the dip value of every system size shown in Fig.~\ref{fig:FermiBoseGap}, for which the results are presented in Fig.~\ref{fig:GapsExtrapolate}. The extrapolated results show a finite single-particle gap $0.055(5)t$ and a vanishing spin gap at the transition point in the thermodynamic limit. Similar results are also obtained for $\lambda/t=0.2$ (as presented and discussed in Appendix~\ref{sec:ResultOf005}) with larger single-particle gaps at the corresponding transition points. Thus, we can conclude that across the TPT the critical mode is bosonic (spin) instead of fermionic, similar to that of free fermion systems. The finite single-particle excitation gap is also consistent with the fact that poles of single-particle Green's function at the transition point are absent.

\section{Summary and discussion}
\label{sec:SumDiscuss}

In this work, we have studied the ground-state properties of the Kane-Mele model with cluster charge interaction in a numerically exact way via large-scale PQMC simulations. We have discovered a topologically nontrivial KVBS phase, as the combination of a QSHI phase with spin Chern number $C_s=-1$ and a long-range KVBS order with spontaneous $Z_3$ symmetry breaking. We have also identified a quantum phase transition from a QSHI phase with $C_s=+1$ to this KVBS phase. Across this transition, the spin excitation gap shows closing and reopening while the single-particle gap remains finite. We have also observed appearance of two zeros of the single-particle Green's function at the transition point, which is consistent with the change of the spin Chern number. Our PQMC results of correlation functions for KVBS order suggest the quantum phase transition to be of first order. 

Our work in this paper explicitly shows {\it a coexistence of  topological symmetry protection and spontaneous symmetry breaking} in interacting fermion systems. We also note that the coexistence of intrinsic topological orders and charge orders has recently be studied and confirmed by exact diagonalization~\cite{Kourtis2014,Kourtis2018} in correlated fermion systems. Combined into a complete piece, these exotic quantum phases and corresponding phase transitions have greatly extended our knowledge of topological phases of matter as well as their quantum phase transitions. As a future study, it will be interesting to investigate whether a spontaneous symmetry breaking quantum phase transition, which is accompanied by a topological phase transition, can be continuous.

\begin{acknowledgments}
Y.Y.H thanks Zi-Xiang Li for valuable discussions. This work was supported by the National Science Foundation of China (Grants No. 11874421 and No. 11774422). R. Q. H. was supported by the Fundamental Research Funds for the Central Universities, and the Research Funds of Renmin University of China (Grant No. 18XNLG11). Computational resources were provided by the Physical Laboratory of High Performance Computing at Renmin University of China and National Supercomputer Center in Guangzhou with Tianhe-2 Supercomputer. The Flatiron Institute is a division of the Simons Foundation.
\end{acknowledgments}

\appendix
\label{sec:ResultOf005}

\begin{figure}[b]
  \includegraphics[width=0.99\columnwidth]{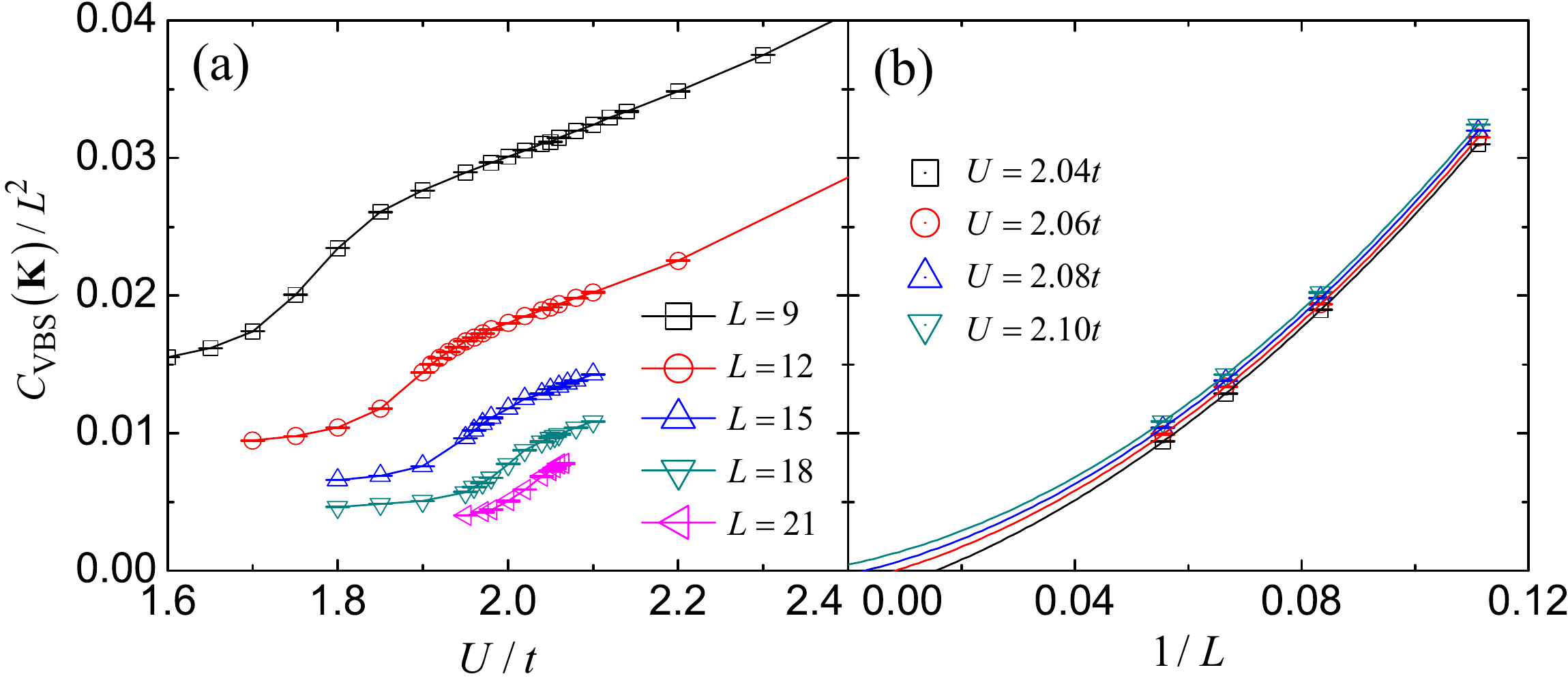}
  \caption{\label{fig:s1} Similar calculations choosing parameter $\lambda=0.05t$ compared with main text. (a) structure factor ${R_B}\left( {U,L} \right)/L^2$. (b) finite-size scaling of structure factors of KVBS order with $L=9,12,15,$ and $18$.}
  \end{figure}

\section{Additional QMC results for $\lambda/t=0.05$ and $\lambda/t=0.2$}
In the main test, we have shown joint PQMC results for $\lambda/t=0.05$ and $\lambda/t=0.2$ with the ground-state $\lambda$-$U$ phase diagram in Fig.~\ref{fig:LatPhDigm}(c). In this appendix, we present the additional QMC results for $\lambda/t=0.05$ and $\lambda/t=0.2$ which are not shown in the main text.

\begin{figure}[b]
  \includegraphics[width=0.99\columnwidth]{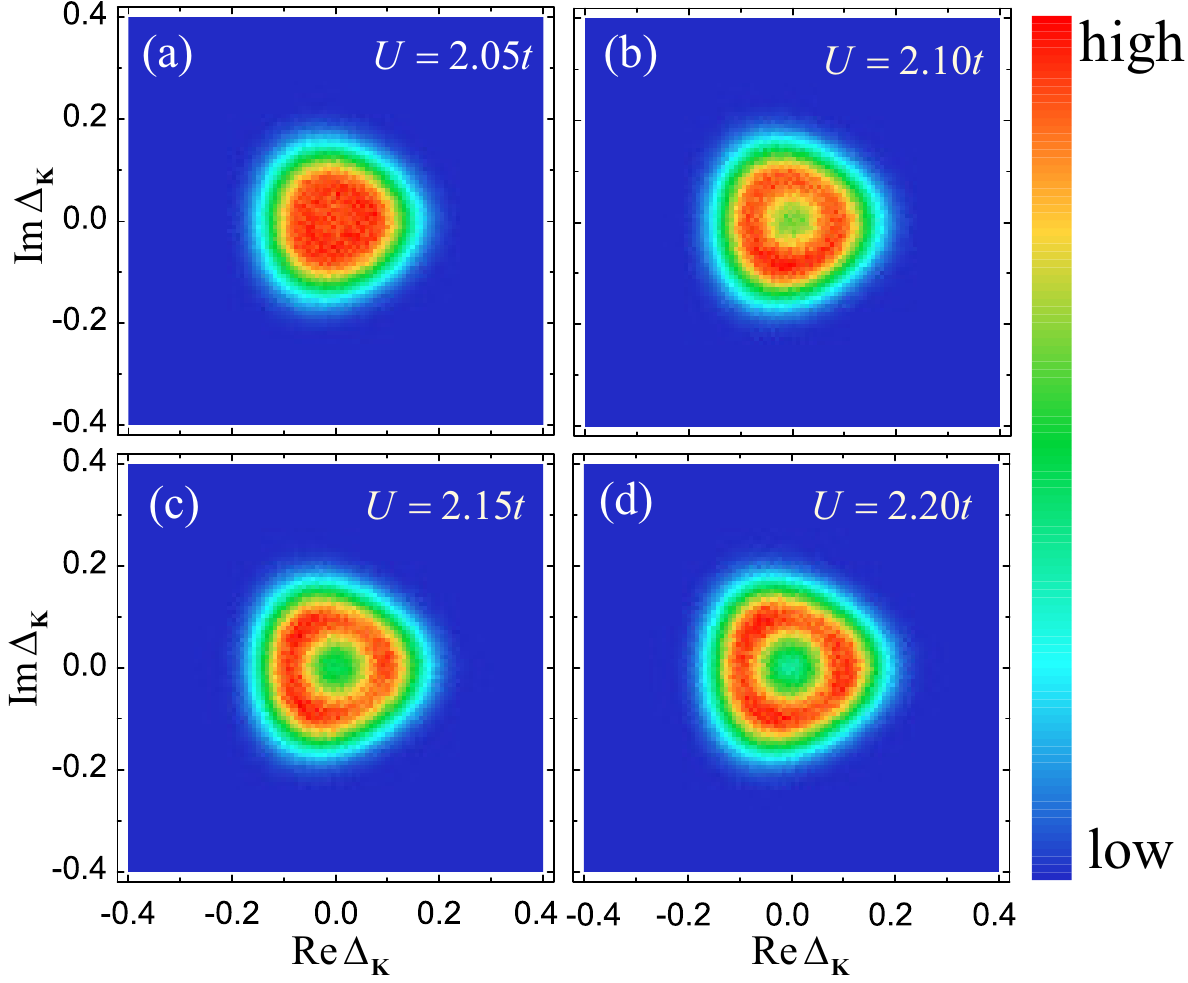}
  \caption{\label{fig:s2} Histogram of KVBS order parameter $\Delta_{\mathbf{K}}$ for $\lambda/t=0.05$ across the phase transition with $L=12$.}
  \end{figure}

\begin{figure}[h]
    \includegraphics[width=0.99\columnwidth]{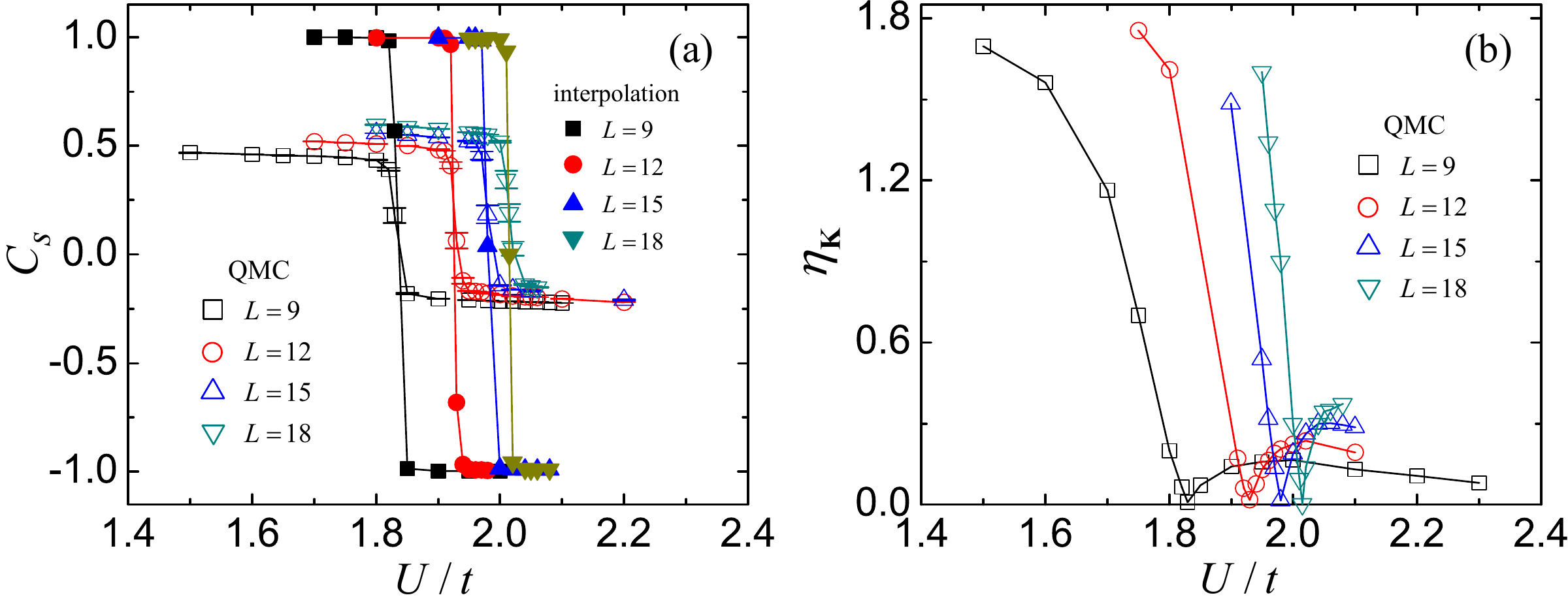}
    \caption{\label{fig:s3} (a) The spin Chern number $C_s$, and (b) positive eigenvalue of $\mathbf{G}_{\sigma}(i\omega=0,\mathbf{K})$ matrix across the TPT for $\lambda/t=0.05$. Interpolation technique is also used here.}
  \end{figure}

For $\lambda/t=0.05$ case, the results of KVBS structure factor, the histogram of KVBS order and the spin Chern number are shown in Figs.~\ref{fig:s1}, ~\ref{fig:s2} and \ref{fig:s3}, respectively. First, from the finite-size scaling of $C_{\rm VBS}(\mathbf{K})/L^2$, we can determine the phase transition point $U_c/t=2.09(1)$, as shown in Fig.~\ref{fig:s1}. From the histogram of $\Delta_{\mathbf{K}}$ presented in Fig.~\ref{fig:s2}, we can clearly observe the evolution of KVBS order, which is absent for $U/t=2.05$ and appears for $U/t=2.15$, in consistence with Fig.~\ref{fig:s1}. On the other side, the spin Chern number $C_s$ as shown in Fig.~\ref{fig:s3}(a) shows finite-size effect, and it's converging to the value of $U_c/t$ obtained from Fig.~\ref{fig:s1}. After the interpolation calculations, we obtain integer quantized values of Cs changing from $C_s=+1$ to $C_s=-1$, suggesting the same TPT as the one for $\lambda/t=0.2$ presented in Fig.~\ref{fig:TopInvtEigVal}. Besides, the zeros of single-particle Green's function also appears at the transition point, which is also consistent with $\lambda/t=0.2$ case. 

As discussed in the main text, our QMC data suggests first-order quantum phase transition from the disordered QSHI phase to the QSHI phase with long-range KVBS order for $\lambda/t=0.2$, especially the results of histogram of $\Delta_{\mathbf{K}}$. However, we are unfortunately not able to draw such a reliable conclusion of the quantum phase transition for $\lambda/t=0.05$ case. We should note that for $\lambda=0$ case, the quantum phase transition from Dirac semimetal to KVBS phase is continuous~\cite{XiaoYan2018}. This means that there might exists finite-size scaling crossover behavior~\cite{Yuanyao2018} when $\lambda$ evolves from zero to $0.2$ (first-order phase transition according to our results). Furthermore, it means that even if the phase transition for $\lambda/t=0.05$ is of first order, significantly larger system sizes (than the ones shown in Fig.~\ref{fig:s1}) are needed to resolve that. However, the results of KVBS structure factor shown in Fig.~\ref{fig:s1} fail to give a good data collapse, which actually contradicts the hypothesis of continuous phase transition. Theoretically, the results of excitation gaps shown in Fig.~\ref{fig:FermiBoseGap} and Fig.~\ref{fig:GapsExtrapolate} explicitly shows the finite single-particle gap, which fits the physical picture discussed in Sec.~\ref{sec:DisOrdTrans} and thus also supports a first order phase transition. 

\begin{figure}[h]
  \includegraphics[width=0.99\columnwidth]{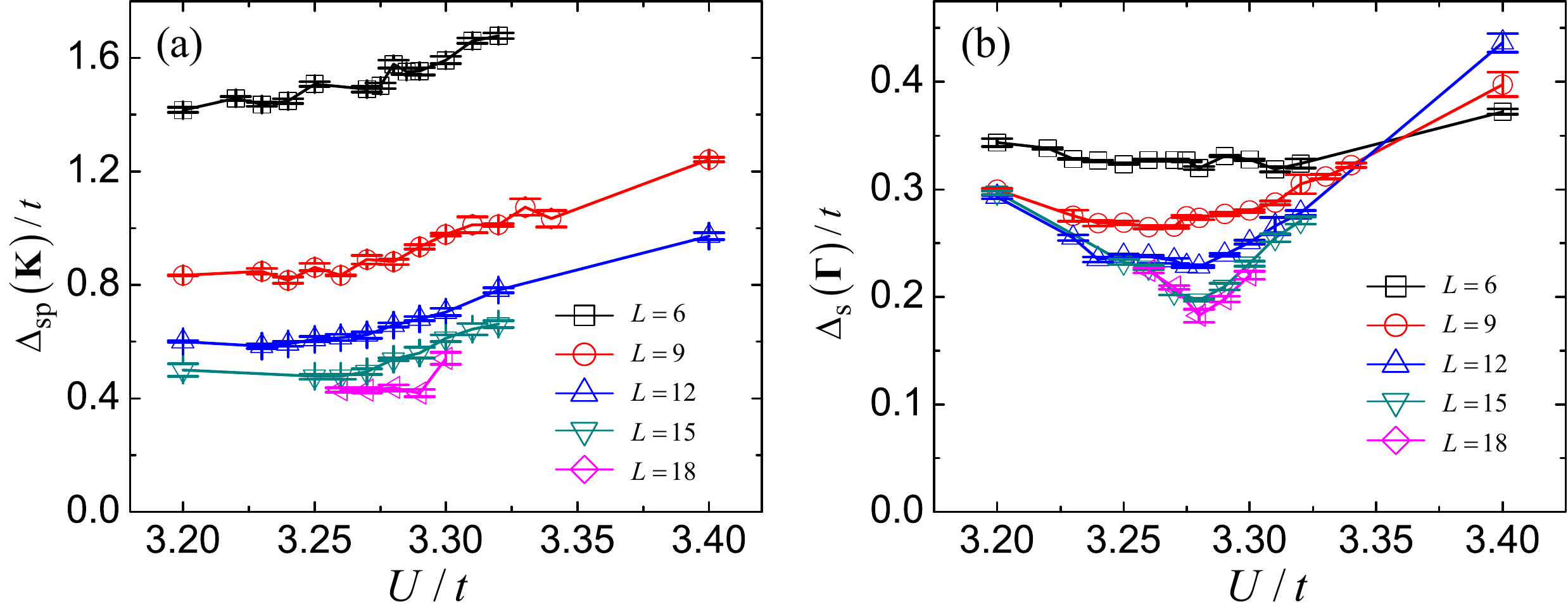}
  \caption{\label{fig:s7} (a) Singel-particle gap and (b) spin gap versus $U/t$ near the phase transition point with $\lambda/t=0.2$.}
\end{figure}

\begin{figure}[h]
  \includegraphics[width=0.7\columnwidth]{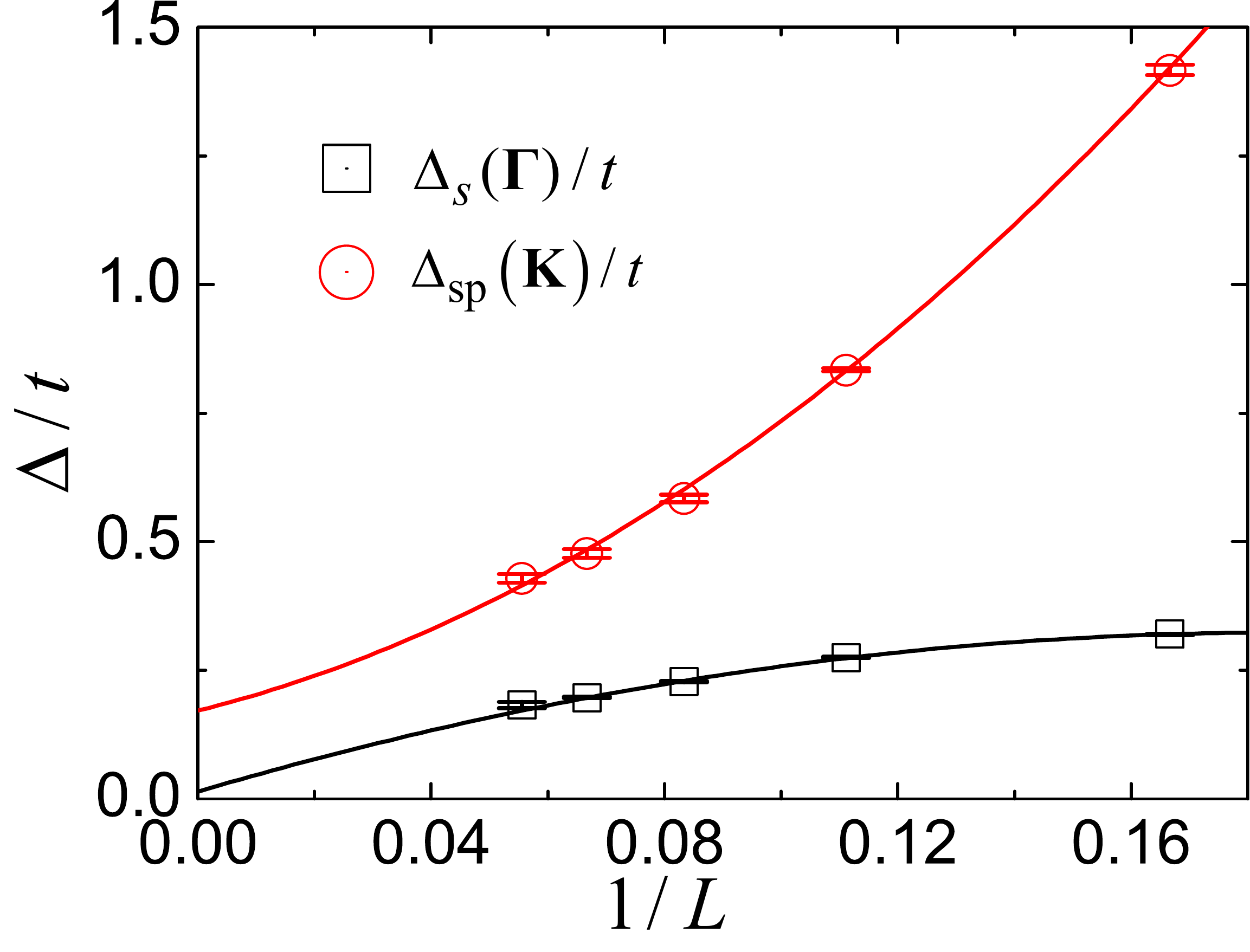}
  \caption{\label{fig:s8} Finite-size scaling of the single-particle gap $\Delta_{sp}(\mathbf{K})/t$ and the spin gap $\Delta_{s}(\boldsymbol{\Gamma})/t$ with the dip values of different system sizes in Fig.\ref{fig:s7} for $\lambda/t=0.2$. Quadratic fittings are applied.}
\end{figure}

For the $\lambda/t=0.2$ case, the results of excitation gaps in fermion and spin channels as well as their finite-size scalings are shown in Figs.~\ref{fig:s7} and ~\ref{fig:s8}. Similar to $\lambda/t=0.05$ case shown in Fig.~\ref{fig:FermiBoseGap}, we can observe that the spin gap is significantly smaller than the single-particle gap. After the extrapolation of the dip values of both gaps to the thermodynamic limit as shown in Fig.~\ref{fig:s8}, the single-particle gap reaches a value close to $0.2t$ while the spin gap vanishes at the transition point. Thus, the spin degrees of freedom becomes critical while the fermion channel is gapped across the phase transition for $\lambda/t=0.2$.

\section{Mean-field results}
\label{sec:MeanField}

\begin{figure}[t]
  \includegraphics[width=0.99\columnwidth]{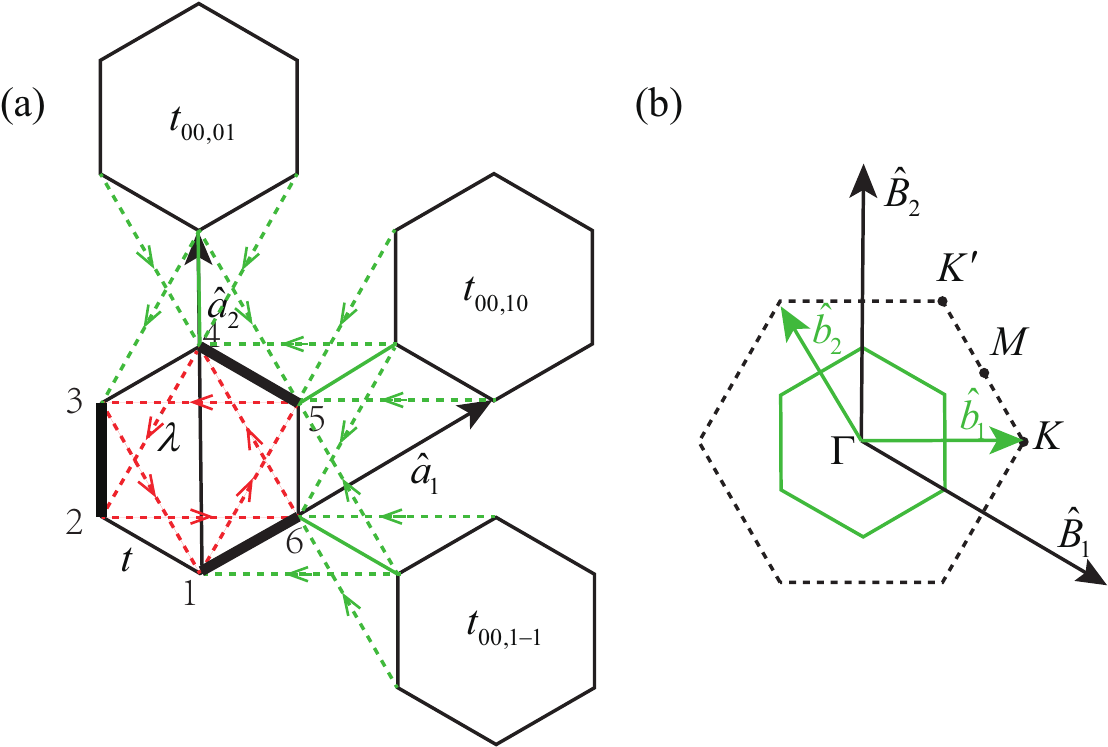}
  \caption{\label{fig:s4} The lattice structure and the Brillouin zone of the mean-field Hamiltonian. (a) Each site in an unit cell is marked with an integer and the lattice vectors are also given as $\hat{a}_1$ and $\hat{a}_2$. The bold lines stand for the electron hoppings that couple to the order parameter $m_{\rm{KVBS}}$. We also draw the three nearest unit cells $t_{{\bf{R}}_i,{\bf{R}}_j}$ which contribute the hopping matrixes $T^{{\bf{R}}_i,{\bf{R}}_j}$ to the Bloch Hmailtonian $H_{\bf{k}}$. (b) The reciprocal lattice vectors $\hat{b}_1$, $\hat{b}_2$ are given corresponding to the lattice vectors $\hat{a}_1$ and $\hat{a}_2$, respectively. The first Brillouin zone is marked by the solid green line. By comparison, We also draw the reciprocal lattice vector and the Brillouin zone of Hamiltonian (\ref{eq:model}) with $\hat{B}_1$, $\hat{B}_2$ and the black dashed line, respectively.}
  \end{figure}

Based on the QMC results presented in Sec.~\ref{sec:DisOrdTrans}, we construct the mean-field Hamiltonian of model (\ref{eq:model}) as follows,
\begin{eqnarray}
\label{eq:mf-hamiltonian}
{H_{{\rm{MF}}}} =&&  -t\sum_{\langle\mathbf{i},\mathbf{j}\rangle\alpha}(c_{\mathbf{i}\alpha}^+c_{\mathbf{j}\alpha} + c_{\mathbf{j}\alpha}^+c_{\mathbf{i}\alpha}) \\ \nonumber
 &&+ i\lambda\sum_{\langle\langle \mathbf{i},\mathbf{j} \rangle\rangle\alpha\beta}\nu_{\mathbf{ij}}(c_{\mathbf{i}\alpha}^+\sigma_{\alpha\beta}^z c_{\mathbf{j}\beta} - c_{\mathbf{j}\beta}^+\sigma_{\beta\alpha}^z c_{\mathbf{i}\alpha} )  \\ \nonumber
 &&- m_{\rm{VBS}}\sum\limits_{{\bf{u}},\sigma} (c_{{\bf{u}}2,\sigma}^+ c_{{\bf{u}}3,\sigma} + c_{{\bf{u}}4,\sigma }^+c_{{\bf{u}}5,\sigma} \\ \nonumber 
&& \hspace{2.5cm} + c_{{\bf{u}}6,\sigma}^ + c_{{\bf{u}}1,\sigma}  + \rm{H.c.}).
\end{eqnarray}
The first two terms are exactly the same as the non-interacting part of Hamiltonian (\ref{eq:model}). We replace the cluster charge interaction term by the third term in Eq.~(\ref{eq:mf-hamiltonian}), which describes the coupling between the KVBS mean-field parameter $m_{\rm VBS}$ and the electron hopping in the bonds marked by the bold line in the inset of Fig.~\ref{fig:LatPhDigm}(c). And the subscripts $\bf{u}$ and $i$ in $c_{{\bf{u}}i,\sigma}^+$ in the third term of Eq.~(\ref{eq:mf-hamiltonian}) are indexes for unit cell and the sites in that unit cell as shown in Fig.~\ref{fig:s4}(a), respectively.

%After the Fourier transform, we can derive the Bloch Hamiltonian $H_{\bf{k}} = T^{00,00} + (T^{00,01} + T^{00,10} + T^{00,0-1} + h.c.)$ at each $\bf{k}$, where $T^{\bf{R}_i,\bf{R}_j}$ are the hopping matrixes contributed from electron hopping between unit cell from $\bf{R}_j$ to $\bf{R}_i$:

\begin{figure}[t]
  \includegraphics[width=0.99\columnwidth]{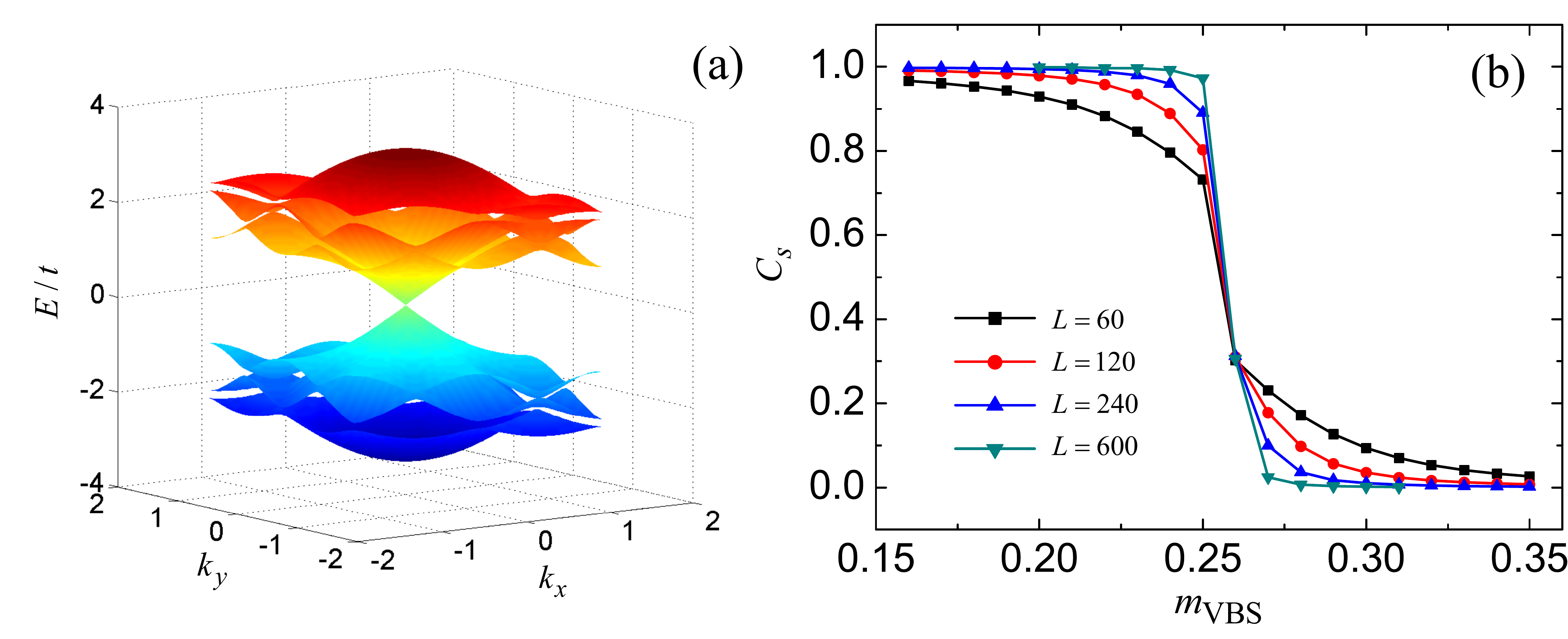}
  \caption{\label{fig:s5} Numerical results for Hamiltonian (\ref{eq:mf-hamiltonian}) with $\lambda/t=0.05$. (a) Energy band in fist Brillouin zone at the gapless point ($\Gamma$) with $m_{\rm{VBS}}/t\sim0.26$. (b) Spin Chern number $C_s$ jump from $+1$ to $0$ before and after the gapless point with $L=60,120,240,$ and $600$, where $L=L_x=L_y$ equals to the number of unit cells along both $x$ and $y$ direction.}
\end{figure}

%\begin{eqnarray}
%\nonumber
%{T^{00,00}} &=& \left({\begin{array}{*{20}{c}}
%  0&{ - t}&{i\lambda }&0&{ - i\lambda }&{ \!-\! t}\\
%  { - t}&0&{  \!-\! t}&{i\lambda }&0&{ - i\lambda }\\
%  { - i\lambda }&{ \!-\! t}&0&{ - t}&{i\lambda }&0\\
%  0&{ - i\lambda }&{ - t}&0&{  \!-\! t}&{i\lambda }\\
%  {i\lambda }&0&{ - i\lambda }&{ \!-\! t}&0&{ - t}\\
%  {\!-\! t}&{i\lambda }&0&{ - i\lambda }&{ - t}&0
%  \end{array}} \right) \\ \nonumber
%  &-& {m_{{\rm{VBS}}}}\left( {\begin{array}{*{20}{c}}
%    0&0&0&0&0&1\\
%    0&0&1&0&0&0\\
%    0&1&0&0&0&0\\
%    0&0&0&0&1&0\\
%    0&0&0&1&0&0\\
%    1&0&0&0&0&0
%    \end{array}} \right) ,\\ \nonumber
%{T^{00,01}} &=& {e^{i{\bf{k}}{{\bf{R}}_{01}}}}\left( {\begin{array}{*{20}{c}}
%  0&0&0&0&0&0\\
%  0&0&0&0&0&0\\
%  { - i\lambda }&0&0&0&0&0\\
%  { - t}&{ - i\lambda }&0&0&0&{i\lambda }\\
%  {i\lambda }&0&0&0&0&0\\
%  0&0&0&0&0&0
%  \end{array}} \right), \\ \nonumber
%{T^{00,10}} &=& {e^{i{\bf{k}}{{\bf{R}}_{10}}}}\left( {\begin{array}{*{20}{c}}
%  0&0&0&0&0&0\\
%  0&0&0&0&0&0\\
%  0&0&0&0&0&0\\
%  0&{ - i\lambda }&0&0&0&0\\
%  {i\lambda }&{ - t}&{ - i\lambda }&0&0&0\\
%  0&{i\lambda }&0&0&0&0
%  \end{array}} \right), \\ \nonumber
%  {T^{00,1 - 1}} &=& {e^{i{\bf{k}}{{\bf{R}}_{1 - 1}}}}\left( {\begin{array}{*{20}{c}}
%  0&0&{i\lambda }&0&0&0\\
%  0&0&0&0&0&0\\
%  0&0&0&0&0&0\\
%  0&0&0&0&0&0\\
%  0&0&{ - i\lambda }&0&0&0\\
%  0&{i\lambda }&{ - t}&{ - i\lambda }&0&0
%  \end{array}} \right).
%\end{eqnarray}

Then, we can solve the mean-field Hamiltonian by diagonalizing the Bloch Hamiltonian at each $\bf{k}$ with different order parameter $m_{\rm{VBS}}$. Setting $\lambda/t=0.05$, our system is simply the Kane-Mele model and is a gapped system with $C_s=+1$ for $m_{\rm{VBS}}=0$ case. With increasing $m_{\rm VBS}$, the system experiences a gap closing and reopening at $\boldsymbol{\Gamma}$ point around $m_{\rm{VBS}}/t\sim0.26$ as shown in Fig.~\ref{fig:s5}(a). At the same time, we find the spin Chern number $C_s$ also jumps from $C_s=+1$ to $0$ as shown in Fig.~\ref{fig:s5}(b).

\section{Raw data of dynamic quantities}
\label{sec:RawdataDynamic}

\begin{figure}[h]
\includegraphics[width=0.99\columnwidth]{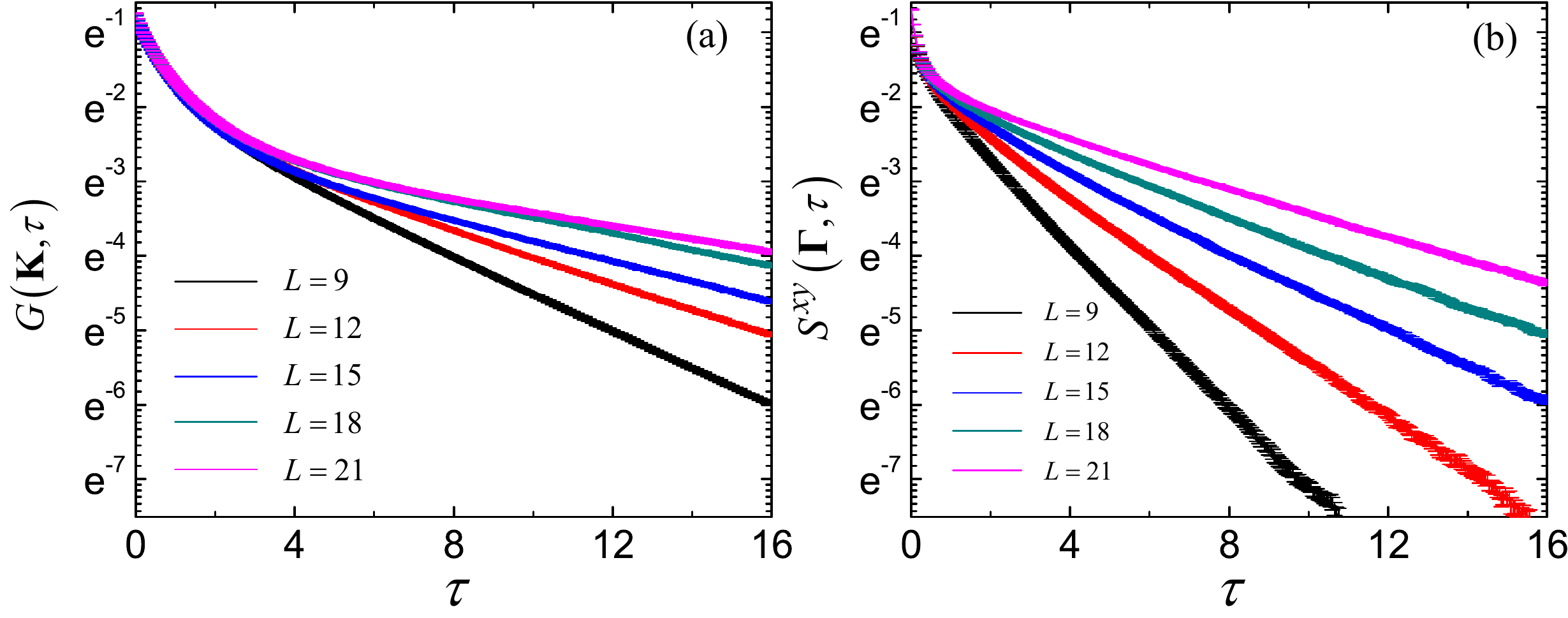}
\caption{\label{fig:RawData} (a) Imaginary-time single-particle Green's function and (b) spin-spin correlation function at the quantum phase transition point $U/t=2$ for $\lambda=0.05t$ with $L=9,12,15,18,$ and $21$.}
\end{figure}

The single particle gap $\Delta_{sp}$ and spin gap $\Delta_s$ shown in Sec.~\ref{sec:FermiBoseGaps} in the main text is obtained from raw data of imaginary-time single-particle Green's function and spin-spin correlation function defined in Eq.~(\ref{eq:Dynamic}). Here we present the raw data of $G(\mathbf{K},\tau)$ and $S^{xy}(\boldsymbol{\Gamma},\tau)$ in Fig.~\ref{fig:RawData}. We have obtained similar results for all other parameters. The raw data with high quality in Fig.~\ref{fig:RawData} allows us to reliably extract the excitation gaps as presented and discussed in Sec.~\ref{sec:FermiBoseGaps}.

\bibliography{myref}

\end{document}